# Effect of isotope disorder on the Raman spectra of cubic boron arsenide

Akash Rai,[1] Sheng Li,[2] Hanlin Wu[2], Bing Lv,[2] David G. Cahill [1]

[1]*Department of Materials Science and Engineering and Materials Research Laboratory, University of Illinois at Urbana-Champaign, Urbana, IL 61801,*

[2]*Department of Physics, University of Texas at Dallas, Richardson, TX 75080, USA*

(Dated: 10 August 2020)

ABSTRACT

Boron arsenide (c-BAs) is at the forefront of research on ultrahigh thermal conductivity materials. We present a Raman scattering study of isotopically tailored cubic boron arsenide single crystals for 11 isotopic compositions spanning the range from nearly pure c-$^{10}$BAs to nearly pure c-$^{11}$BAs. Our results provide insights on the effects of strong mass disorder on optical phonons and the appearance of two-mode behavior in the Raman spectra of mixed crystals. Strong isotope disorder also relaxes the one-phonon Raman selection rules, resulting in disorder-activated Raman scattering by acoustic phonons.

INTRODUCTION

In 2013, first-principle calculations by Lindsay *et al*. predicted a remarkable room temperature thermal conductivity ($\kappa$) for cubic boron arsenide (c-BAs) of 2200 W/m-K, comparable to that of diamond [1]. This calculation [1] considered only lowest order anharmonicity. The high room temperature $\kappa$ was attributed to significantly suppressed three-phonon scattering in c-BAs due to a combination of a large frequency gap between acoustic and optical modes (a-o gap), bunching of acoustic branches, and relatively small optical phonon bandwidth [1]. More recent computational studies that include the next higher-order term in the anharmonicity found that significant four-phonon scattering in c-BAs reduces the predicted thermal conductivity at room temperature to 1400 W/m-K [2,3]. The experimental value is approximately 1000 W/m-K [4,5,6].

In this work, we examine another extreme property of the lattice dynamics of c-BAs: the effect of strong isotope disorder on the vibrational modes cannot be described by the weak-scattering limit of the coherent potential approximation that is typically successful in predicting the frequencies and lifetimes of optical phonons in isotopically modified crystals [7,8]. The strong deviation from virtual crystal behavior in c-BAs was first identified in a 2013 study of the vibrational Raman spectra for natural isotope abundance and $^{11}$B isotopically-enriched crystals [9]. Here, we extend that work to a study of eleven c-BAs isotopically mixed crystals with isotope compositions that span the range from nearly pure c-$^{10}$BAs to nearly pure c-$^{11}$BAs. The large number of compositions allows us to systematically examine how the Raman spectra evolve as a function of mass disorder.

Within the harmonic approximation, the Raman frequency shift ($\nu$) of the zone center optical phonon of a homogenous crystal scales inversely as the square root of the reduced mass ($\mu$) of the two atoms that make up the basis of atoms in a zinc blende or wurtzite crystal structure:

$$\nu \propto \mu^{-1/2} \qquad (1)$$

Within the weak-scattering limit of the coherent potential approximation (CPA), this scaling law can be modified to account for isotope disorder and anharmonicity through a phonon self-energy term. This phonon self-energy has real and imaginary parts related to the phonon frequency shift and spectral broadening, respectively [8]. Phonon self-energies due to isotopic disorder have been extensively studied [7,8,9,10,11]. Phonons in a homogeneous crystal have infinite lifetime in the harmonic approximation. In real crystal with no isotopic disorder, the phonon lifetime is finite due to anharmonicity and the phonon lineshape is Lorentzian. The Lorentzian lineshape is slightly distorted in the presence of isotope disorder.

A comprehensive Raman study by Cardona *et al*. described the effect of isotopic disorder on the lattice dynamics of several elemental and compound semiconductors [8]. They calculated the self-energies due to isotopic disorder and obtained a detailed picture of the phonon renormalization by simulating lineshape properties within the framework of the CPA. More recent theoretical calculations by Mahan derived the effect of atomic isotopes on the phonon modes of a crystal and concluded that in solving for the vibrational modes, mixing of modes by isotope scattering is an important process [12].

The vibrational properties of cubic boron arsenide with natural isotope abundance ($^{nat}$BAs) and cubic boron arsenide isotopically enriched with $^{11}$B ($^{11}$BAs) were studied by Hajidev *et al.* using Raman spectroscopy and first principle density function perturbation theory (DFPT) [9]. In $^{nat}$BAs, they observed an anomalous two-mode behavior not seen in other isotopically disordered crystals; they tentatively attributed this behavior to the localization of $^{10}$B vibrational states. They also observed a lack of LO-TO splitting in the measured Raman spectrum of $^{11}$BAs and further investigated the small LO-TO splitting in $^{nat}$BAs using DFPT [9]. The calculated phonon dispersion of c-BAs includes overbending of the upmost optical phonon branch [9]; i.e., the frequency of phonons on the non-degenerate optical branch is higher than that of the zone-center phonon, similar to the form of the optical phonon dispersion of diamond [13]. They evaluated the real part of phonon self-energy in $^{nat}$BAs arising due to isotopic disorder at zone center as $\Delta_{dis} \approx$ -3 cm$^{-1}$, with small disorder-induced broadening $\Gamma_{dis} \approx$ 0.4 cm$^{-1}$ [9].

The two-mode behavior seen in the Raman spectra of $^{nat}$BAs is often observed in mixed crystals of the type $LM_{1-x}N_x$, where the Raman frequencies of the LM and LN components differ significantly [14,18]. At any intermediate compositions $x$, two sets of frequencies are often observed; one set of frequencies is related to the LO and TO modes of the lighter LM component, and the other set of frequencies is related to the LO and TO modes of the heavier LN component [18]. There have been numerous attempts to provide a theoretical criterion for predicting one-or two-mode behavior in mixed diatomic crystal. These criteria are extensively reviewed by Sen and Hartmann [16].

Chang and Mitra proposed a modified-random-element-isodisplacement (MREI) model that assumed that all cation-anion pairs vibrate in phase, as in the $k = 0$ optic mode of a perfect diatomic crystal [14]. This leads to a criterion for two-mode behavior depending only on the masses of constituents,

$$m_M^{-1} > m_L^{-1} + m_N^{-1}. \tag{2}$$

Here, $\mu_{LN}^{-1} = m_L^{-1} + m_N^{-1}$ is the reduced mass of the compound LN.

This criterion states that a mixed crystal that displays two-mode behavior has a substituting atom with a mass smaller than the reduced mass of the compound formed by the other two atoms. They successfully applied this criterion to predict one- and two-mode behavior in various mixed crystals [14]. Figure 1 shows the summary of this condition tested on mixed crystals by Chang and Mitra together with c-$^{10}$B$_{1-x}$$^{11}$B$_x$As, i.e., a $^{10}$B impurity in $^{11}$BAs. The dashed line represents the condition stated in Eq. 2 and forms a separating boundary between one- and two-mode behavior in mixed crystals. As seen in Figure 1, c-BAs lies almost exactly on the dashed line and is therefore at the cusp of satisfying this condition of two-mode behavior.

EXPERIMENTAL

We synthesized single crystals of isotopically tailored c-BAs by a modified chemical vapor transport (CVT) method using $I_2$ as the transport agent [5]. The B source materials, $^{10}$B powder (Alfa Aesar, 99.9% chemical purity and >96 atomic% $^{10}$B ), and $^{11}$B powder (Alfa Aesar, 99.9% chemical purity and >96 atomic% $^{11}$B), are mixed with the molar ratio of $^{10}$B: $^{11}$B = $x$ : 1-$x$ ($x$ varies from 0 to 1 at intervals of 0.1). The B powder is mixed with As chunks (Alfa Aesar, 99.999%) with the atomic ratio of B:As =1:7. The mixture of B and As source materials is then sealed in an evacuated quartz tube together with a small amount (typically 1.2 mg cm$^{-3}$ of the volume of the container) of the transport agent ($I_2$). The end of the quartz tube containing the source materials was placed at the high temperature zone of a horizontal furnace at 900°C. The cold zone of the quartz tube was placed at the lower temperature zone of the furnace at 800°C. Crystal growth proceeds over the course of 3 weeks. The synthesized single crystals of c-BAs have a lateral dimension on the order of 0.5 mm and growth facets parallel to the {111} planes.

The isotopic composition of the c-BAs crystals deviates from the nominal composition likely due to the differences of transfer rate during CVT process caused by the sizes and morphologies of the $^{10}$B and $^{11}$B powders and is determined using time-of-flight secondary ion mass spectroscopy (TOF-SIMS). Before measurement, we clean the c-BAs single crystals by ultrasonication in ethanol followed by distilled water for three minutes each. We then ion polish a {111} growth facet for 5 minutes using a GATAN PECS II ion polishing system. The operating parameters of the ion polishing system are an Ar ion energy of 6 keV, ion gun current of 45 µA, and grazing incidence angle of 3°.

Raman spectroscopy measurements are performed using custom-built optical system based on an Acton Insight spectrometer (Princeton Instruments). The excitation wavelength is 488 nm from a Spectra-Physics Cyan laser. The excitation laser is focused by a 20×, N.A.=0.4, objective lens. The laser power on the sample is ≈ 6 mW and the $1/e^2$ intensity radius of the focused laser beam is ≈5 µm. The backscattered Raman light is collected through the same objective and dispersed by a 1200 mm$^{-1}$ grating. The spectrometer is calibrated using a Si wafer. The frequency shift of the Raman active optical phonon in Si

at room temperature is 520 cm$^{-1}$ [19]. We assume a linear relationship between pixel number and Raman shift. We verified our calibration by measuring Raman scattering by O$_2$ molecules in the air adjacent to an Al-coated sample. We measure the O$_2$ stretching vibrational mode frequency as 1555 cm$^{-1}$, within 0.1% of the expected value of 1556.2 cm$^{-1}$ [20].

The expected measured line-shape of a Raman peak is the convolution of a Lorentzian phonon line-shape with the Gaussian response function of the spectrometer. The convolution of a Lorentzian with a Gaussian is a Voigt profile [21]. An approximation for the relation between the full-width-at-half-maximum (FWHM) widths of the Voigt ($f_v$), Gaussian ($f_G$), and Lorentzian ($f_L$) profiles is given by [22]:

$$f_v \approx 0.535 f_L + \sqrt{0.217 f_L^2 + f_G^2} \qquad (3)$$

The FWHM of a Voigt profile fit to experimental data for the Raman peak of Si in our Raman spectrometer is 5.4 cm$^{-1}$. The Voigt profile can be deconvoluted to find the width of the Gaussian spectrometer response function using the previously determined Lorentzian lineshape of Si. The intrinsic FWHM of the Si Raman line is 2.6 cm$^{-1}$ at room temperature [23, 24]. The Gaussian FWHM due to instrumental broadening alone is 3.9 cm$^{-1}$.

RESULTS AND DISCUSSION

Figure 2 shows the normalized Raman spectra of isotopically tailored cubic boron arsenide single crystals from 650 cm$^{-1}$ to 750 cm$^{-1}$ with $^{11}$B-rich (panel (a)) and $^{10}$B-rich (panel (b)) compositions. The Raman spectrum for nearly pure $^{11}$BAs has a single symmetric peak corresponding to the Raman active longitudinal optical (LO) vibrational mode at zone center. We label the peak that we associate with $^{11}$B-related vibrational modes as P1. The Voigt profile fit to P1 for nearly pure $^{11}$BAs (x=0.01), is shown in Figure 3(a). After deconvolution by the instrument resolution function, the Lorentzian FWHM of this peak is $1.2 \pm 0.2$ cm$^{-1}$ corresponding to an optical phonon lifetime of $4.4 \pm 0.6$ ps [25].

As isotope disorder increases in $^{11}$B-rich crystals, see Figure 2(a), the P1 peak broadens and a second peak, labelled P2, appears where the phonon density-of-states of $^{11}$BAs and $^{10}$BAs overlap [9]. To fit the spectra using symmetric Voigt profiles, however, we must introduce a third peak (P3) at frequencies between P1 and P2, at intermediate compositions. We are unsure of the physical origin of P3 and P3 may be an artifact created by our assumption that the P1 and P2 line-shapes are symmetric. This third peak P3 is seen for all $^{11}$B-rich isotopically mixed crystals except $x$=0.01. In $^{10}$B-rich disordered crystals ($0.5 < x < 0.97$), however, peak P3 is not seen when the measured spectra are fit to multiple peaks. This is because P1 and P2 are relatively close in frequency(< 10 cm$^{-1}$ apart) and including peak P3 leads to overfitting of the measured spectra.

In the previous study of natural isotope abundance c-BAs by Hajidev *et al.*, the P2 peak was attributed to localized $^{10}$B-related optical modes [9]. A recent study on localization of phonons in mass-disordered alloys have shown that for binary isotopic disorder, light impurities can induce localized modes beyond the bandwidth of the host system [26]. The observed two mode behavior is also consistent with the Chang and Mitra criterion discussed above (Eq. 2) [14].

The Raman spectrum for nearly pure $^{10}$BAs crystals has a single symmetric P2 peak, corresponding to the Raman active LO optical phonons at zone center. The Voigt profile fit to P2, after deconvolution by

the instrumental resolutions, yields a Lorentzian FWHM of $1.5 \pm 0.3$ cm$^{-1}$. This linewidth corresponds to an optical phonon lifetime of $3.6 \pm 0.5$ ps [25]. As $x$ decreases from 0.97 to 0.56, i.e., as $^{11}$B is substituted for $^{10}$B, the P2 peak broadens and a P1 peak appears. We conclude that the two-mode behavior observed in $^{nat}$BAs by Hajidev *et al.* [9] is observed for all isotopic compositions of c-$^{10}$B$_x$$^{11}$B$_{1-x}$As except the compositions that are nearly isotopically pure, $x = 0.01$ and $x = 0.97$.

Figure 4 summarizes the positions and widths of peaks P1 and P2 as a function of isotopic composition. The dashed line in Figure 4(a) shows the expected mean-field dependence of the optical phonon frequency on the composition through the reduced mass (Eq. 1). The trend in the Raman shift of the peaks P1 and P2 as a function of isotopic composition is reminiscent of the two-mode behavior observed in mixed crystals that satisfy the Chang and Mitra criteria [14]. Since the LO-TO splitting is smaller than our instrument resolution, at any intermediate compositions $x$, only two frequencies are observed; one frequency is related to the optical vibrational modes of the lighter $^{10}$B-like vibrations (peak P2), and the other frequency is related to the optical vibrational modes of the heavier $^{11}$B-like vibrations (peak P1).

Within the framework of the weak scattering limit of the coherent potential approximation (CPA), the spectral broadening of the Raman peak of an isotopically disordered crystal is due to the decrease in the phonon lifetime caused by mass disorder; the broadening is expected to increase with the increasing isotopic mass disorder and reach a maximum at $x = 0.5$. Hajidev *et al*. calculated the disorder-induced broadening withing the CPA and found an extremely small broadening of 0.4 cm$^{-1}$ [9]. For isotopically disordered c-BAs crystals, however, the observed broadening is nearly two orders of magnitude larger and therefore cannot easily be attributed to changes in optical phonon lifetime. The Lorentzian FWHM of both peak P1 and P2 is estimated by fitting the data to a symmetric Voigt profile and deconvoluting the Voigt profile with the instrumental resolution function. The Lorentzian FWHM of P1 and P2 derived from this procedure is summarized in Figure 4(b).

We also observe disorder-activated Raman scattering where the usual crystal momentum selection rules are partially relaxed and phonon frequencies with a high density of states appear in the one-phonon Raman spectra. We plot the low frequency region of the Raman spectra as Figure 5(a). In isotopically disordered sample with compositions $x = 0.24, 0.36, 0.60$ and $0.74$, we observe peaks corresponding to TA (L) at 132 cm$^{-1}$, TA (W) at 218 cm$^{-1}$ and LA (W) at 275 cm$^{-1}$. The frequencies of the acoustic phonons do not shift significantly with B isotopic composition and therefore we expect these disorder-activated to be independent of isotopic composition. The high density of states at these frequencies are apparent in the phonon density of states calculated by density function perturbation theory and plotted in Figure 5(b). We also observe a weak peak in the data for all compositions at 394 cm$^{-1}$, corresponding to two phonon Raman scattering from transverse acoustic phonons with crystal momentum near the X point.

Finally, we discuss the effects of strong isotopic disorder on the high frequency regime of the Raman spectra that includes Raman scattering by combinations of two optical phonons, see Figure 6(a). We compare our Raman scattering data to calculations of the optical phonon density of states in Figure 6(b) where we have plotted the p-DOS against twice the phonon frequency. The second order spectra of the $^{11}$B-enriched crystal ($x = 0.01$) shows two-phonon Raman scattering peaks corresponding to contribution from 2TO(X) and 2TO(W) at 1267 cm$^{-1}$, 2TO(L) at 1348 cm$^{-1}$ and 2LO($\Gamma$) at 1405 cm$^{-1}$. These two-phonon peaks are less prominent in the $^{10}$B-enriched crystal ($x = 0.97$) due to a larger background in the Raman intensities that we attribute to Raman scattering by free carriers. The experimentally observed shift in wavevector at $\Gamma$ point of $^{10}$B-enriched crystal ($x=0.97$) compared to $^{11}$B-enriched crystal ($x = 0.01$) is $\approx 52$ cm$^{-1}$, smaller but comparable to the calculated shift in the phonon density of states with the reduced mass of the crystal $\approx 65$ cm$^{-1}$.

CONCLUSION

In summary we have studied the effect of isotope disorder on the vibrational spectra of isotopically tailored c-BAs single crystals using Raman spectroscopy at room temperature. The two mode behavior that is predicted by the Chang and Mitra criterion for the substitution of $^{10}$BAs in $^{11}$BAs is mostly clearly observed for $^{11}$BAs-rich compositions. For $^{10}$BAs-rich compositions, the two-mode behavior is not well developed because the spacing in frequency between the two modes is of the same order as their width.

ACKNOWLEDGMENT

The authors thank Prof. N.K. Ravichandran and Prof. D.A. Broido for providing the DFPT calculated phonon density of state data for c-$^{11}$BAs and c-$^{10}$BAs at 0 K. This work was supported by the ONR MURI grant N00014-16-1-2436, and ONR grant N00014-19-1-2061.

FIGURES

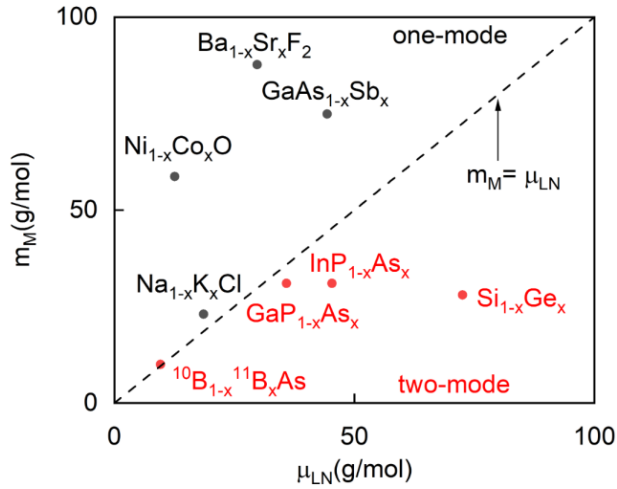

figure 1: Summary of the two-mode behavior criterion (Eq. 2) proposed by Chang and Mitra,[14] tested on selected mixed crystals of the type $LM_{1-x}N_x$ (mixture of the compound LM and the compound LN) along with c-$^{10}B_{1-x}^{11}B_xAs$ ($^{10}B$ substituting in $^{11}BAs$). The y axis is the mass of the substituting element M ($m_M$) and the x axis is the reduced mass ($\mu_{LN}$) of the compound LN. The dashed line represents the condition $m_M = \mu_{LN}$ and forms a separating boundary between one- and two-mode crystals. The behavior of mixed crystals is drawn from Refs. [14].

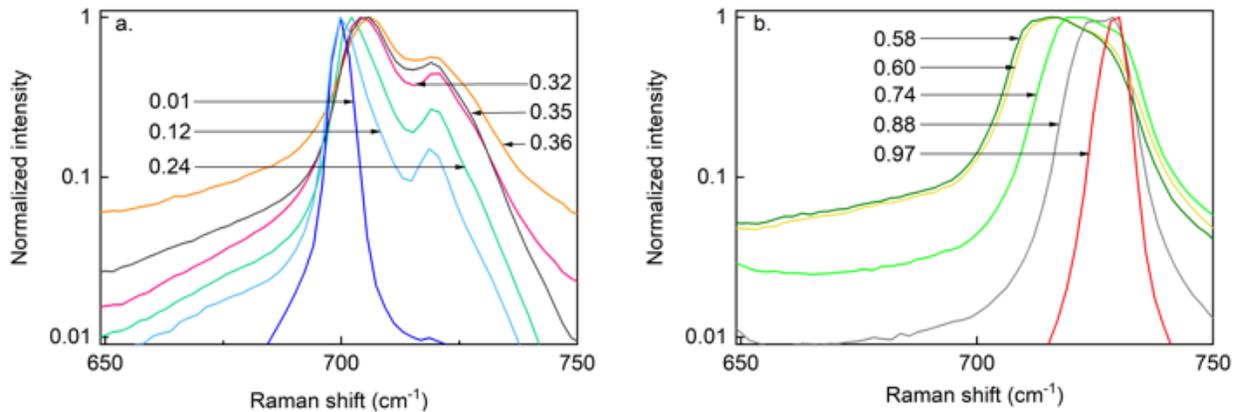

Figure 2: Raman spectra of isotopic mixed crystals of $^{10}B_x^{11}B_{1-x}As$ for (a) $0 < x < 0.4$; and (b) $0.5 < x < 1.0$. The spectra are measured at room temperature with 488 nm laser excitation on {111} growth facets in a backscattering geometry. The measured Raman intensity is normalized by the maximum intensity of the spectrum for each sample. The y-axis is logarithmic.

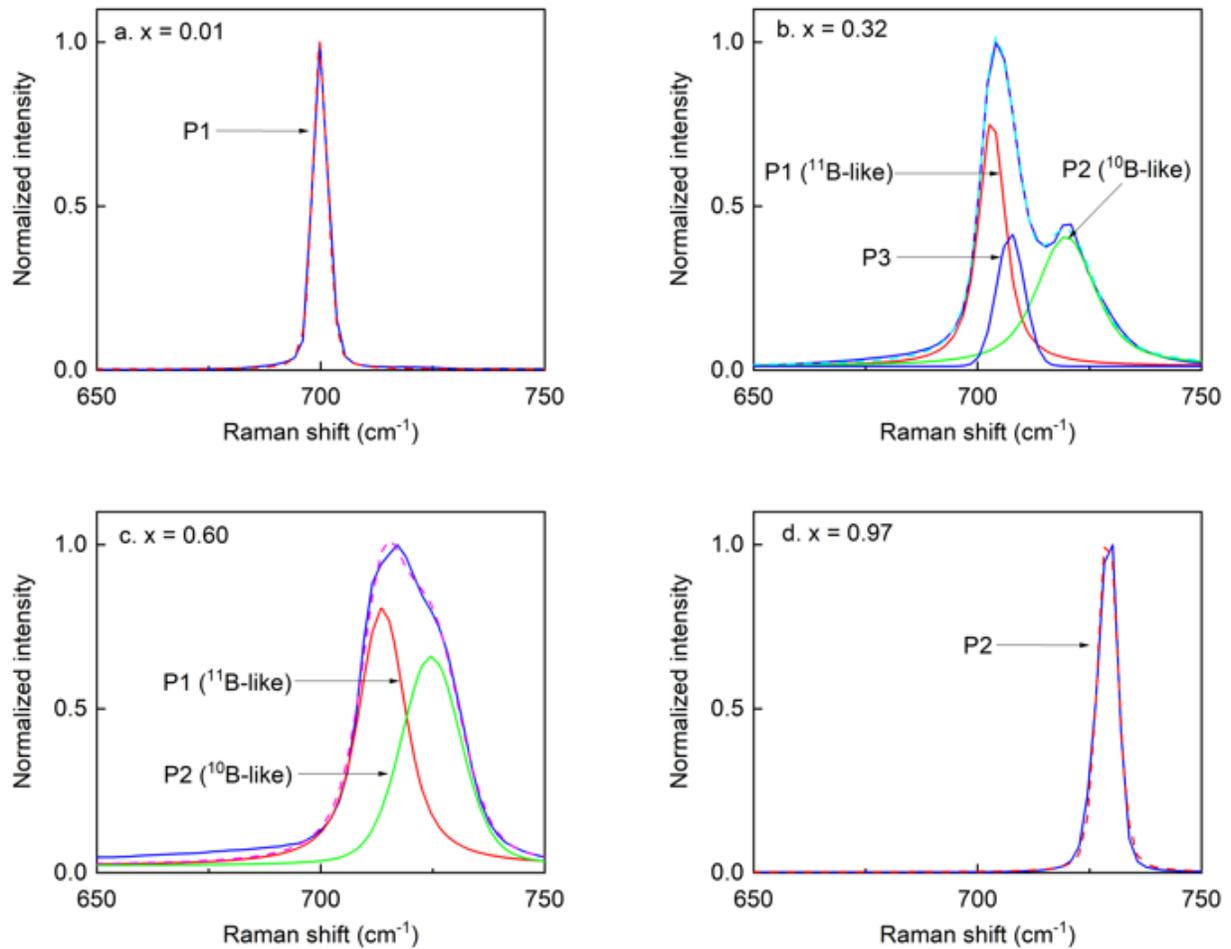

Figure 3: Raman spectra of compositions (a) $x = 0.01$ (b) $x = 0.32$ (c) $x = 0.60$ (d) $x = 0.97$. The experimental data are shown as solid lines and fits to a symmetric Voigt profile are shown as dashed lines. The measured spectra are fit to multiple peaks. We attribute peaks P1 and P2 to $^{11}$B-related and $^{10}$B-related vibrational modes, respectively. In the case of intermediate compositions, e.g., $x=0.32$, we include a third peak P3 in the fitting procedure to account for Raman scattering intensity that is not described by P1 and P2.

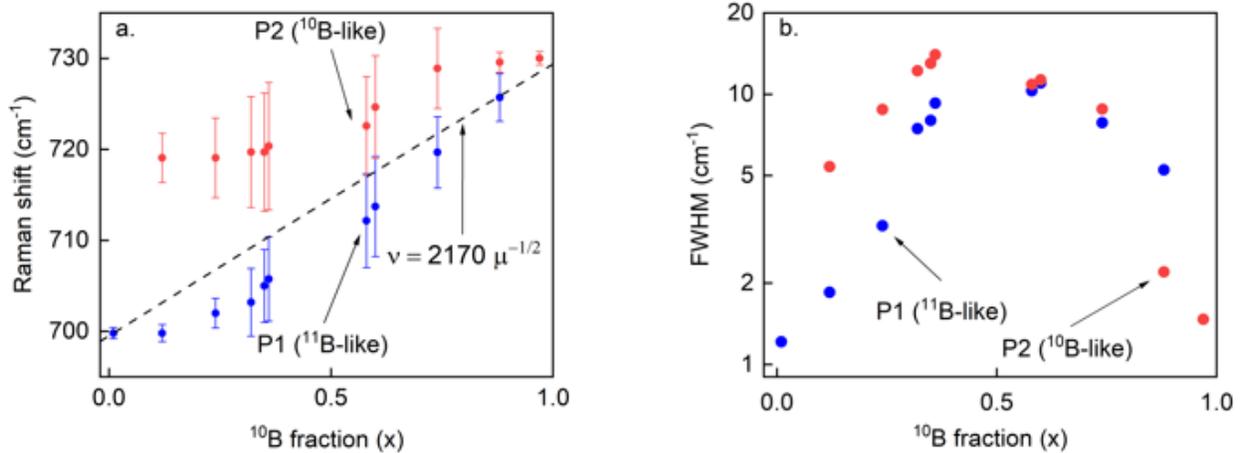

Figure 4: (a) The measured Raman shift ($\nu$) corresponding to peaks P1 (blue data points) and P2 (red data points), due to $^{11}$B-like and $^{10}$B-like optical vibrational modes respectively, with respect to $^{10}$B isotopic composition ($x$). The dashed line indicates the expected scaling of the zone center optical phonon frequency with the reduced mass ($\mu$), see Eq. 1. For each isotopic composition, the Lorentzian and Gaussian FWHM of P1 and P2 can be estimated by fitting the experimental data with Voigt profile fit and deconvoluting by the Gaussian instrumental broadening. In panel (a), the Lorentzian FWHM of P1 and P2 are depicted as error bars surrounding the peak Raman shift for each composition. In panel (b), the Lorentzian FWHM of P1 (blue data points) and P2 (red data points) are plotted as a function of $^{10}$B composition ($x$). The y-axis of panel (b) is logarithmic.

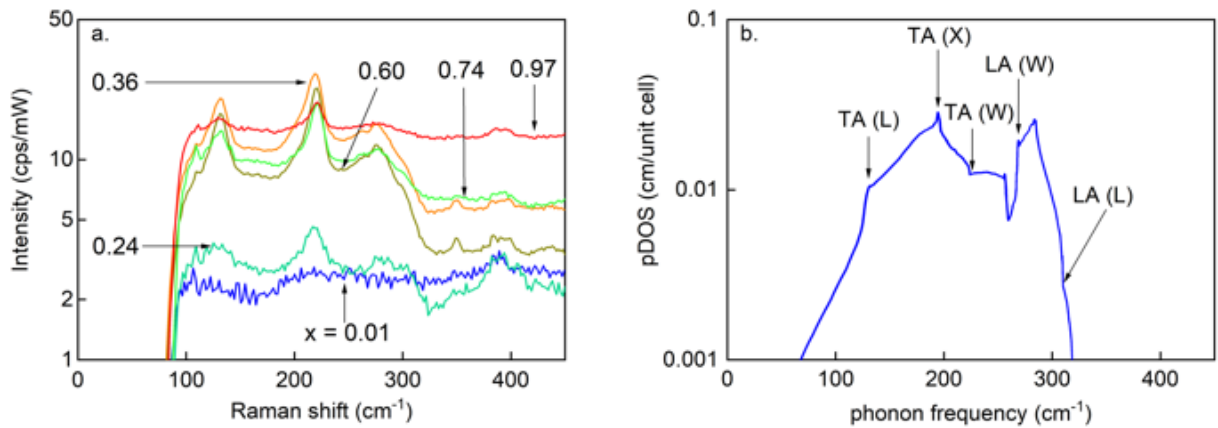

Figure 5: (a) Measured Raman spectra of low frequency acoustic phonons in c-$^{10}$B$_x$$^{11}$B$_{1-x}$As for various $^{10}$B compositions (x). The intensity in counts per second is normalized by the power of the 488 nm laser excitation in mW units. (b) Calculated phonon density of states (p-DOS) showing critical points associated with transverse acoustic (TA) and longitudinal acoustic (LA) phonons at points on the Brillouin zone boundaries, W, X, and L. Isotopically tailored crystals show disordered induced Raman scattering from acoustic phonons which are not seen in the isotopically enriched crystal (x=0.01 and 0.97). The most prominent features are scattering by the TA phonons at the L and X points. The y-axes of both panels (a) and (b) are logarithmic.

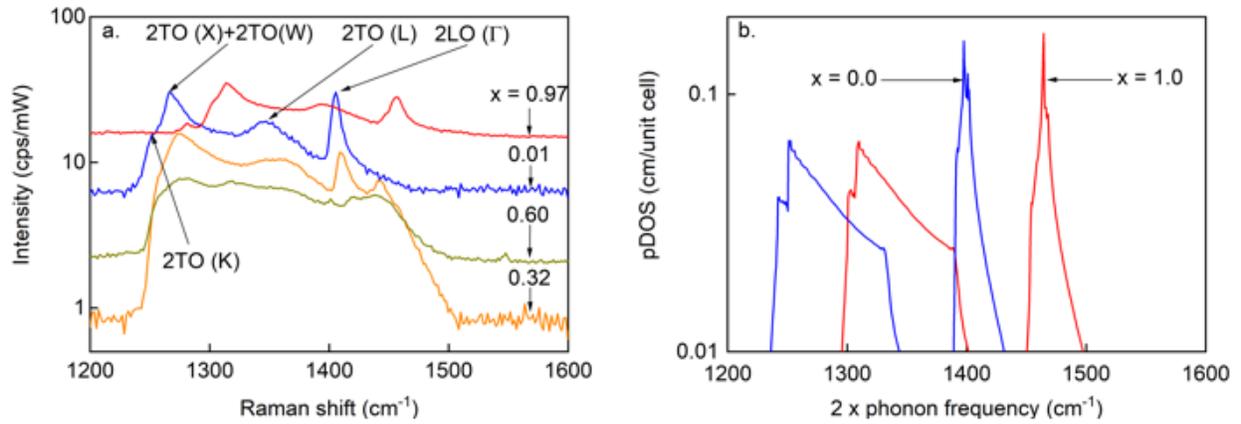

Figure 6: (a) Measured Raman spectra in the high frequency regime that involves Raman scattering from combinations of two optical phonons. The intensity in counts per second is normalized by the power of the 488 nm laser excitation in mW units. (b) Optical phonon density of states (pDOS) calculated by density functional perturbation theory plotted against twice the phonon frequency to facilitate comparison with the experiment data in panel (a).